\documentclass[priprint, amsmath,amssymb,
nobibnotes,nofootinbib]{revtex4}
\usepackage{soul}
\usepackage{aas_macros}
\usepackage{graphicx}
\usepackage[caption=false]{subfig}
\usepackage{newtxtext,newtxmath}
\usepackage{ae,aecompl}
\usepackage{hyperref}
\usepackage{graphicx}
\usepackage{epsf}
\usepackage{longtable}
\usepackage{rotating}
\usepackage{bm}
\usepackage{color}
\usepackage{amsmath}

\usepackage{booktabs}
\usepackage{braket}
\usepackage{multirow}
\setcitestyle{square}
\usepackage{xcolor}

\begin{document}

\title{Constraining Neutrino Mass with the Tomographic Weak Lensing Bispectrum}
\author{William R. Coulton}
\affiliation{Institute of Astronomy and Kavli Institute for Cosmology Cambridge, Madingley Road, Cambridge, CB3 0HA, UK}
\affiliation{Joseph Henry Laboratories, Princeton University, Princeton, NJ 08544, USA}
\author{Jia Liu}
\affiliation{Department of Astrophysical Sciences, Princeton University, Peyton Hall, Princeton, NJ 08544, USA}
\author{Mathew S. Madhavacheril}
\affiliation{Department of Astrophysical Sciences, Princeton University, Peyton Hall, Princeton, NJ 08544, USA}
\author{Vanessa B\"ohm}
\affiliation{Berkeley Center for Cosmological Physics, University of California, Berkeley, CA 94720, USA}
\affiliation{Lawrence Berkeley National Laboratory, 1 Cyclotron Road, Berkeley, CA 93720, USA}
\author{David N. Spergel}
\affiliation{Department of Astrophysical Sciences, Princeton University,Peyton Hall, Princeton, NJ 08544, USA}
\affiliation{Center for Computational Astrophysics, Flatiron Institute,162 5th Avenue, 10010, New York, NY, USA}

\begin{abstract}
We explore the effect of massive neutrinos on the weak lensing shear bispectrum using the Cosmological Massive Neutrino Simulations \cite{Liu2018MassiveNuS:Simulations}. We find that the primary effect of massive neutrinos is to suppress the amplitude of the bispectrum with limited effect on the bispectrum shape. The suppression of the bispectrum amplitude is a factor of two greater than the suppression of the small scale power-spectrum. For an LSST-like weak lensing survey that observes half of the sky with five tomographic redshift bins, we explore  the constraining power of the bispectrum on three cosmological parameters: the sum of the neutrino mass $\sum m_\nu$, the matter density $\Omega_m$ and the amplitude of primordial fluctuations $A_s$. Bispectrum measurements alone provide similar constraints to the power-spectrum measurements and combining the two probes leads to significant improvements than using the latter alone. We find that the joint constraints tighten the power spectrum $95\%$ constraints by $\sim 32\%$  for $\sum m_\nu$, $13\%$ for $\Omega_m$ and $57\%$ for $A_s$ .
\end{abstract}
\maketitle
\section{Introduction}
The neutrino mass sum ($\sum m_\nu$) is now confirmed to be non-zero via the discovery of oscillations between the flavor eigenstates~\cite{Becker-Szendy1992,Fukuda1998,Ahmed2004}. While we now know the differences between the squared masses of the three neutrino species, their absolute mass sum remains unknown. Cosmological surveys have the potential to constrain the mass sum of neutrinos by measuring their impact on the expansion history and growth of structure in the universe~\cite{LesgourguesPastor2006,Wong2011,Yu2018TowardsInformation}. In the early universe, massive neutrinos of mass $\le$ a few eV behave like radiation and hence delay the onset of linear growth. At late times, due to their large thermal velocities, neutrinos can stream out of the cold dark matter~(CDM) potential wells and hence suppress the growth of structure below the free-streaming scale. 

Neutrino oscillation data imply a lower limit of $\approx$~0.06eV~\cite{Olive:2016xmw}. Current surveys are already approaching this limit. The Planck team obtained an upper limit of $\sum m_\nu< 0.12$~eV (95\% CL)~\cite{PlanckCollaboration2018PlanckParameters}, combining measurements of the cosmic microwave background (CMB) temperature, low-$\ell$ polarization, CMB lensing, and baryon acoustic oscillation
measurements~\cite{Beutler2011,Anderson2014,Ross2015}. Constraining the neutrino mass sum is also one of the major missions of several upcoming large cosmological surveys, including the Large Synoptic Survey Telescope~(LSST)\footnote{\url{http://www.lsst.org} }, Wide-Field Infrared Survey Telescope~(WFIRST)\footnote{\url{http://wfirst.gsfc.nasa.gov} }, Euclid\footnote{\url{http://sci.esa.int/euclid} }, Simons Observatory\footnote{\url{http://www.simonsobservatory.org} }, and CMB-S4\footnote{\url{https://cmb-s4.org}}. 

The effect of massive neutrinos is most important in the quasi-linear and nonlinear regime. For $\approx 0.1$~eV neutrinos in a standard Planck cosmology, for example, most effects due to massive neutrinos are at scales smaller than the free-streaming length, which is around 110 Mpc today. Two-point statistics, while being the most common analysis tool for present-day cosmological surveys, can not capture the full information content in the nonlinear regime. Therefore, in recent years there has been increasing interest in statistics beyond the two point function. Combining two point statistics with higher order statistics, such as Minkowski functionals \citep{shirasakiyoshida2014, Petri2015EmulatingFunctionals}, clipping transforms \citep{Giblin2018,Simpson2016}, peak counts~\citep{Liu2015CosmologyData,Liux2015,Kacprzak2016CosmologyData,Martinet2017,Shan2017} and bispectrum measurements \citep{Kayo2013InformationMatrix,Takada2004}, has been shown to tighten cosmological parameters by a factor of $\approx$2. 

In this work, we focus on the weak lensing bispectrum, which is the harmonic equivalent of the three point function. Weak lensing signals of the CMB and galaxies are detected through measurements of the distortion of the background source (in the case of CMB, the temperature or polarization anisotropies, while in the case of galaxies, the galaxy shapes). They are promising probes of the underlying matter density field. The bispectrum is the lowest order perturbative correction to the Gaussian distribution, and hence is zero for a purely Gaussian field. As the weak lensing signal is an integrated quantity, we expect deviations from Gaussianity to be relatively small and thus expect the bispectrum to contain most of the non-Gaussian information in the lensing field. Since the first three-point shear measurements were made \citep{Bernardeau2002,Jarvis2004}, there has been significant work to understand how systematics impact bispectrum measurements such that, now, unbiased measurements of cosmological parameters can be performed with the bispectrum \citep{Semboloni2011WeakStatistics, Fu2014CFHTLenS:Correlations}.  Complementary to the shear bispectrum, there has been work on the CMB lensing bispectrum by \citet{Namikawa2016CMBStructure,Pratten2016ImpactCMB,Bohm2016BiasStructure}. Most relevant to this work is the work by \citet{Namikawa2016CMBStructure} that found  combining CMB lensing bispectrum measurements with power-spectrum measurements would  provide a $\sim 30\%$ improvement on constraints on the dark energy equation of state $w$ and $\sum m_\nu$ (however this work did not include the post-Born corrections which will likely alter this). Whilst previous work on the weak lensing bispectrum has focused on constraining the six parameters of $\Lambda$CDM and models of dark energy \citep{Sato2013ImpactEstimation}, we explore the constraints on $\sum m_\nu$.

Throughout this work we use a flat cosmology with the following parameters: Hubble parameter $h=0.7$, primordial scalar spectrum power-law index $n_s=0.97$, baryon density $\Omega_b=0.046$, and the dark energy equation of state $w=1$. We vary the sum of the neutrino masses $\sum m_\nu$, the matter density $\Omega_m$ and the amplitude of fluctuations $A_s$.

This paper is organized in the following manner: in Section \ref{sec:methods} we overview the simulations used and the analysis methods. In Sections \ref{sec:powerspec} and \ref{sec:bispec} we describe the main effects of neutrino mass on the power spectrum and bispectrum. In Section \ref{sec:constraints} we present the constraints obtain from power spectrum, bispectrum and their combination and we then conclude in Section \ref{sec:conclusions}. In Appendix \ref{app:powspecTests} we present a set of robustness tests of our method.



\section{Methodology}\label{sec:methods}
\subsection{Simulations}\label{sec:massiveNuSims}
The Cosmological Massive Neutrino Simulations~(\texttt{MassiveNuS})~\cite{Liu2018MassiveNuS:Simulations} consist of a large suite of 101 N-body simulations\footnote{The \texttt{MassiveNuS} data products, including galaxy and CMB lensing convergence maps, N-body snapshots, halo catalogues, and merger trees, are publicly available at \url{http://ColumbiaLensing.org}.}, with three varying parameters $\Sigma m_\nu$, $A_s$, and $\Omega_m$. The simulations use the public code \texttt{Gadget-2}~\cite{springel2005}, with a box size of 512~Mpc$h^{-1}$ and 1024$^3$ CDM particles, accurately capturing structure growth at $k<$10~$h$~Mpc$^{-1}$. \texttt{MassiveNuS} adopts a fast linear response algorithm~\citep{AB2013,Bird2018}, where neutrinos are treated using linear perturbation theory and their clustering is sourced by the full nonlinear matter density\footnote{The neutrino patch \texttt{kspace-neutrinos} is publicly available at \url{https://github.com/sbird/kspace-neutrinos}}. This method avoids the shot noise and high computational costs that are usually associated with particle neutrino simulations. The code has been tested robustly and agreements with particle neutrino simulations are found to be within 0.2\% for $\Sigma m_\nu < 0.6$ eV.

Galaxy convergence maps are generated with the ray-tracing code \texttt{LensTools}~ \citep{Petri2016Lenstools}\footnote{\url{https://pypi.python.org/pypi/lenstools/}}. The N-body snapshots are first cut into 4 planes, each with comoving thickness 126 Mpc$h^{-1}$. 4096$^2$ regularly spaced light rays from the center of the z = 0 plane are then shot backwards in redshift, spreading over a 3.5$^2$ deg$^2$ solid angle, and their trajectories are tracked until the source planes at $z$=0.5, 1.0, 1.5, 2.0, 2.5 (five galaxy source planes). This  ray-tracing calculation does not assume the Born approximation and thus automatically includes the post-Born terms \cite{Dodelson2005}. In total, through randomly rotating and shifting lens planes, 10,000 convergence map realizations are generated per cosmological model per redshift, each with map size 3.5$^2$ deg$^2$ and 512$^2$ pixels. For each realization, the maps at different source redshifts are ray-traced through the same large scale structure and hence are properly correlated. Whilst our box is relatively small we believe it is sufficiently large to capture the effects of super sample variance \citep{Takada2013PowerCovariance} as, due to our small patch size, we are able to include modes that are a factor of $\sim 4$ times larger than the map.

To model the covariance matrices, we also generate an additional set of simulations at the fiducial model ($\Sigma m_\nu=0$, $A_s=2.1\times10^{-9}$, and $\Omega_m=0.3$), with different initial conditions. This is necessary to avoid the correlation between the model and the covariance noises during likelihood estimation, which can artificially underestimate the error size \citep{Carron2013OnMatrices}. We use 90,000 simulated convergence maps from five N-body simulations, where we use the method described in \citet{Sato2009SIMULATIONSEFFECTS} to draw large numbers of different convergence maps from a set of N-body simulations. The validity of this approach was explored in \citet{Petri2016SampleRequired}. 

\subsection{Convergence maps}
As described in Section \ref{sec:massiveNuSims} we ray-trace through our simulations to generate convergence maps. The convergence map is a weighted projection of the matter density field
\begin{align}\label{eq:defKappa}
\kappa^{X}(\bm{\theta})=\int \mathrm{d}z W^{X}(z)\delta(\chi(z)\bm{\theta},z),
\end{align}
where $\delta(\chi(z)\bm{\theta},z)$ is the total matter over density at redshift $z$, $\chi(z)$ is the comoving distance, $\bm{\theta}$ is the position on the sky, $W^{X}(z)$ is the lensing kernel for the sources in the $X \in \{0.5,1.0,1.5,2.0,2.5\}$  redshift bin. In this work we assumed that all the sources are distributed delta function planes in redshift, thus the lensing kernels are 
\begin{align}
W^{X}(z)=\frac{3}{2}\Omega_m H_{0}^2\frac{(1+z)\chi(z)}{H(z)c}\frac{\chi(z_X)-\chi(z)}{\chi(z_X)},
\end{align}
where $\Omega_m$ is the present day fractional matter density, $c$ is the speed of light, $H(z)$ is the Hubble's parameter and $z_X$ is the redshift of the source plane. Note that Eq. \ref{eq:defKappa} assumes the Born-approximation but this is not assumed in our simulations, as described in  Section \ref{sec:massiveNuSims}.

As our simulated maps are small, 3.5$^2$ deg$^2$, we will work in the flat-sky approximation and we use the extended Limber approximation \citep{Loverde2008ExtendedApproximation} for calculating theoretical quantities. Thus we can decompose the convergence maps into Fourier coefficients
\begin{align}
\kappa^{X}(\bm{\theta})=\int \frac{\mathrm{d}^2\ell}{4\pi^2} \kappa^{X}_{\mathbf{\ell}} e^{i \bm{\theta}\cdot \bm{\ell}}.
\end{align}

\begin{table}

\centering
\begin{tabular}{l l l l l l }
\hline
 & \multicolumn{5}{ c }{Redshift }\\
\hline
$z_s$ & 0.5 & 1.0 &1.5 & 2.0 & 2.5\\
\hline
$\bar{n}_{\mathrm{gal}}$ & 8.83&13.25&11.15&7.36&4.26\\
\hline
\end{tabular}
\caption{The projected source counts per arcmin$^2$ used in this work. }
\label{tab:nGalPerZ}

\end{table}
Our simulations give us the true $\kappa$ field, however when performing observations we only have access to noisy reconstructions.  For the galaxy-shear convergence maps, the noise arises due to the discrete sampling of galaxies and the intrinsic shape noise of the observed galaxies. In this work we simulate noise levels appropriate for the LSST experiment by adding Gaussian noise to our maps with variance $\sigma_s^2/\bar{n}^{X}_{\rm gal}$, where $\sigma_s=0.3$ is the intrinsic shape noise and $\bar{n}^X_{\rm gal}$ is the projected number of sources per arcmin$^2$. The source counts used in this work are shown in Table \ref{tab:nGalPerZ} and were chosen to be consistent with the projected levels for LSST \citep{LSSTscienceBook}. 

\subsection{Binned Power-spectrum Estimator}
In the flat-sky regime, the power-spectrum between convergence maps at redshifts $X_1$ and $X_2$, $C^{X_1,X_2}_{\ell}$, is defined as
\begin{align}
\langle \kappa^{X_1}_{\bm{\ell_1}}\kappa^{X_2}_{\bm{\ell_2}} \rangle = (2\pi)^2\delta^{(2)}(\bm{\ell_1}+\bm{\ell_2}) C^{X_1,X_2}_{\ell_1}.
\end{align}
where $\delta^{(2)}(...)$ is a 2D Dirac delta function. We estimate the binned power-spectrum as
\begin{align}
\hat{C}^{X_1,X_2}_{i}=\frac{1}{N_i} \int_{\ell_i}^{\ell_{i+1}}\mathrm{d}^2\ell \, \kappa^{X_1}_{\bm{\ell}}{\kappa^{X_2}_{\bm{\ell}}}^*
\end{align}
where $\hat{C}^{X_1,X_2}_{i}$ is the estimated binned power-spectrum for the $i^{th}$ bin, $\ell_i$ ($\ell_{i+1}$) is the lower (upper) boundary of the $i^{th}$ bin, and $N_i$ is the number of modes in each bin. We have pixelated convergence maps and so we discretize the above integrals and replace them with sums. For the power-spectrum analysis we used bins of width $\Delta \ell$=100, a minimum $\ell_{\rm min}$=150 and a maximum $\ell_{\rm max}$=3150.

\subsection{Binned Bispectrum Estimator}
The calculation of the full bispectrum for modern surveys is computationally prohibitive and instead compression methods are used. In this work we use a binned bispectrum estimator based on the work of \citet{Bucher2010,Bucher2016} but adapted to the flat-sky regime. Recall that, under the conditions of homogeneity and isotropy, the bispectrum can be written as \citep{Hu2000WeakApproach,Spergel1999MicrowaveFormalism}
\begin{align}
\langle \kappa^{X_1}_{\bm{\ell_1}}\kappa^{X_2}_{\bm{\ell_2}}\kappa^{X_3}_{\bm{\ell_3}}  \rangle = (2\pi)^2\delta^{(2)}(\bm{\ell_1}+\bm{\ell_2}+\bm{\ell_3}) b^{X_1,X_2,X_3}_{\ell_1,\ell_2,\ell_3}
\end{align}
where $b^{X_1,X_2,X_3}_{\ell_1,\ell_2,\ell_3}$ is the reduced bispectrum between three convergence maps at redshifts $X_1$, $X_2$ and $X_3$. The reduced bispectrum only depends on the magnitude of the $\ell$ modes. The binned bispectrum is then defined as
\begin{align}
\hat{b}^{X_1,X_2,X_3}_{i,j,k}=\frac{1}{N_{i,j,k}} \int_{\ell_i}^{\ell_{i+1}} \frac{\mathrm{d}^2\ell_1 }{(2\pi)^2}
\int_{\ell_j}^{\ell_{j+1}} \frac{\mathrm{d}^2\ell_2 }{(2\pi)^2}
\int_{\ell_k}^{\ell_{k+1}} \frac{\mathrm{d}^2\ell_3 }{(2\pi)^2} (2\pi)^2\delta^{(2)}(\bm{\ell_1}+\bm{\ell_2}+\bm{\ell_3}) \kappa^{X_1}_{\bm{\ell_1}}\kappa^{X_2}_{\bm{\ell_2}}\kappa^{X_3}_{\bm{\ell_3}}
\end{align}
where $N_{i,j,k}$ is number of triplets in each bin, $\ell_i$ ($\ell_{i+1}$) is the lower (upper) boundary of the $i^{th}$ bin and $\hat{b}^{X_1,X_2,X_3}_{i,j,k}$ is the estimated binned bispectrum. The binned bispectrum can be efficiently implemented by first creating a set of filtered maps
\begin{align}
\mathcal{W}^{X}_i(\bm{\theta})=\int_{\ell_i}^{\ell_{i+1}} \frac{\mathrm{d}^2\ell }{(2\pi)^2}\kappa^{X}_{\bm{\ell}} e^{i\bm{\ell}\cdot\bm{\theta}}.
\end{align}
Then the binned bispectrum estimator is reduced to a simple sum over products of filtered maps
\begin{align}
\hat{b}^{X_1,X_2,X_3}_{i,j,k}=\frac{1}{N_{i,j,k}}\int \mathrm{d}^2\theta \,\mathcal{W}^{X_1}_{i}(\bm{\theta})\mathcal{W}^{X_2}_{j}(\bm{\theta})\mathcal{W}^{X_3}_{k}(\bm{\theta}).
\end{align}
As in power-spectrum case for our pixelated maps we replace the integrals with sums and we implement the Fourier transforms with the FFTW3 library \citep{FFTW05}. For the bispectrum analysis we used bins of width $\Delta \ell$=300, a minimum $\ell_{\rm min}$=150, and a maximum $\ell_{\rm max}$=3150. The bin width is chosen to ensure there are closed triangles within our binned configurations. The minimum $\ell$ is governed by the size of patch and the maximum $\ell$ was chosen to immunize our results to the impact of baryonic physics, which is not account for in our simulations.

\subsection{Interpolation}\label{sec:interpolation}
To generate parameter constraints we need to evaluate the likelihoods at points in parameter space beyond the 101 cosmologies we have simulated. Following the work of 
\citet{Heitmann2009THESPECTRUM,Heitmann2013THESPECTRUM}, we build an emulator that interpolates the power-spectrum and bispectrum measurements from the simulated cosmologies. To do this, we use a Gaussian Process with a triaxial squared exponential kernel  \cite
{Rasmussen2006GaussianLearning}. We used the scikit-learn implementation of this algorithm.\citep{Pedregosa2011Scikit-learn:Python}
Gaussian processes are very useful for interpolating from multidimensional scattered data and, as tested in our companion paper by Li et al. (in prep.), the interpolation accuracy is typically within half a percent and significantly smaller than the LSST error bars. In Appendix \ref{app:powspecTests}, we test the robustness of our interpolation methods.


\subsection{Likelihoods}\label{subsec:likelihood}

We used Gaussian likelihoods to describe the distribution of the binned power-spectrum, bispectrum and joint measurements. Thus
\begin{align}\label{eq:likelihoods}
\ln \mathcal{L} \propto \frac{1}{2}\sum_{i,j,X,Y} \left(\hat{S}^{X}_i-\bar{S}^{X}_i \right){\Sigma_{S}^{-1}}^{X,Y}_{i,j} \left(\hat{S}^{Y}_j-\bar{S}^{Y}_j \right)
\end{align}
where $\hat{S}^{X}_i$ is a vector of either the power spectrum, bispectrum or joint measurements for the redshift configuration X and the $i^{th}$ bin configuration, barred quantities denote the mean of the observable, and $\Sigma_{S}$ is the covariances of the statistic. Many of our bins have large numbers of bispectrum triplets, so we expect the resulting distribution relatively Gaussian due to the central limit theorem.  We examined the probability distribution function (PDF) of single bins of the binned-bispectrum and verify that the Gaussian likelihood is a reasonable approximation. 
In addition, as the dominant source of non-Gaussianity in lensing likelihoods comes from the presence of a few rare massive clusters, for experiments that cover very large areas of the sky, like LSST, many different clusters contribute to the maps so that these rare events are less important. 
Other approaches such as the Approximate Bayesian Computation and likelihood-free inference methods may further improve the parameter estimation\citep{Alsing2018MassiveCosmology}.

We assume cosmology-independent covariance matrices and calculate them from a separate set of simulations, as described in Section \ref{sec:massiveNuSims}. 
We remove the bias in the inverse covariance matrix with the correction factor \citep{Hartlap2007WhyMatrix} 
\begin{align}
\Sigma_{\textrm{unbiased}}^{-1}=\frac{n-1}{n-p-2} \Sigma_{\textrm{sample}}^{-1}
\end{align}
where $\Sigma_{\textrm{sample}}^{-1}$ is the inverse of the covariance matrix from the simulations, $n$ is the number of simulations and $p$ is the number of parameters. Our simulated maps have an area of $12.25$ deg$^2$ and so to generate constraints for LSST, which will cover approximately half the sky, we scale our covariance by $f_{\rm sky}^{\rm sim}/f_{\rm sky}^{\rm LSST}=12.25/20626.5$.

We use flat priors that uniformly weight values within the region covered by our simulations and give zero probability to cosmologies outside. Thus
\begin{align}
    P\left(\sum m_\nu \right)= &
\begin{cases}
    \text{const},& \text{if } 0\, \mathrm{eV} \leq \sum m_\nu \leq 0.62\,\mathrm{eV} \\
    0,              & \text{otherwise}
\end{cases}\\
    P\left(\Omega_m\right)=&
\begin{cases}
    \text{const},& \text{if } 0.18\, \leq \Omega_m \leq 0.42 \,  \\
    0,              & \text{otherwise}
\end{cases}\\
    P\left(A_s\right)=&
\begin{cases}
    \text{const},& \text{if } 1.29 \, \leq A_s < 2.91 \,\\
    0,              & \text{otherwise}.
\end{cases}
\end{align}
Finally we use the Markov Chain Monte Carlo (MCMC) with the emcee \citep{Foreman-Mackey2013EmceeHammer} package to sample from the parameter posteriors and the corner package \citep{corner} to display the results.

\section{The effect of neutrino mass on the power-spectrum}\label{sec:powerspec}
\begin{figure}
  \centering
    \includegraphics[width=.60\textwidth]{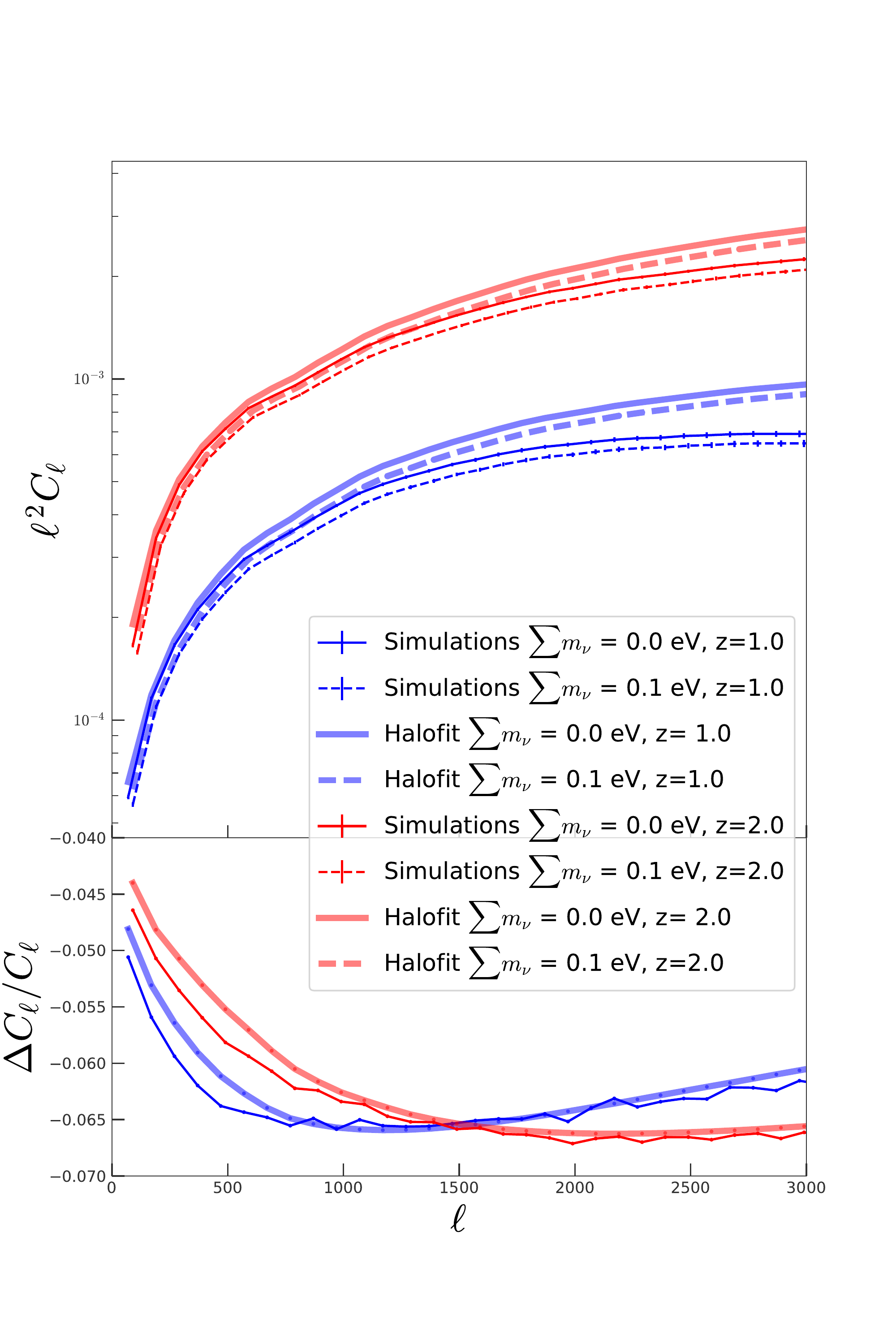}
\caption{
The shear power spectrum for two redshift configurations is plotted for the case of massless neutrinos (solid lines) and the case with $\sum m_\nu=0.1$ eV (dotted lines). We compare the power-spectrum from simulations to that calculated using the HALOFIT matter power spectrum, integrated using the extended Limber approximation \citep{Loverde2008ExtendedApproximation}. The bottom plots show the fractional difference between the massive and massless neutrino power-spectra. The error bars show how well these configurations can be measured with an LSST-like experiment that observes half of the sky. As is discussed in \citet{Liu2018MassiveNuS:Simulations}, the difference between HALOFIT and the simulations at high $\ell$ arises from the finite resolution of the simulations.}
\label{fig:powerspectrumResponse}
\end{figure}

Before presenting the bispectrum results we briefly review the effect of neutrinos on the power-spectrum. For a detailed review of their effects see \citet{Lesgourgues2006MassiveCosmology}. On the largest scales, massive neutrinos behave like cold dark matter,  whereas on scales smaller than $k_{\rm fs} = 0.0072 (\sum m_\nu/0.1eV)^{\frac{1}{2}} (\Omega_m/0.315)^{\frac{1}{2}}h$  Mpc$^{-1}$ they free stream. As can be seen in Fig. \ref{fig:powerspectrumResponse} the main effect of neutrino mass on the power-spectrum is to suppress the amplitude, by $\sim 6 \%$ for $0.1$ eV neutrinos. The suppression shows a scale dependence, which arises as massive neutrinos contribute to the energy density but do not cluster on small scales.  Massive neutrinos also affect the growth of structure and thus power-spectra at different redshifts are affected differently. This effect can be seen in Fig. \ref{fig:powerspectrumResponse} and is explored further in Section \ref{sec:constraints}. The error bars in Fig. \ref{fig:powerspectrumResponse} show the expected uncertainties for an LSST-like experiment; note that while at large scales they are very small, these points are correlated. Finally, in Fig. \ref{fig:powerspectrumResponse} we plot the predictions of HALOFIT\citep{Takahashi2012} using CAMB for the shear power-spectrum. We can see these power-spectra agree well with the results from the simulations. The  simulation's power spectra show a slight lack of power at small scales, which arises due to the resolution limit of the simulations. The effect of neutrino mass on the largest scales is greater in the simulations, which is thought to arise due to the finite box size. On small scales we see slight differences between the effect of neutrinos in the simulations and in the HALOFIT model and this is understood to arise from uncertainties in the HALOFIT model. For a more detailed comparison see \citet{Liu2018MassiveNuS:Simulations}.
%

\section{The effect of neutrino mass on the bispectrum}\label{sec:bispec}

\begin{figure}
\subfloat[Squeezed Configuration]{
  \centering
    \includegraphics[width=.45\textwidth]{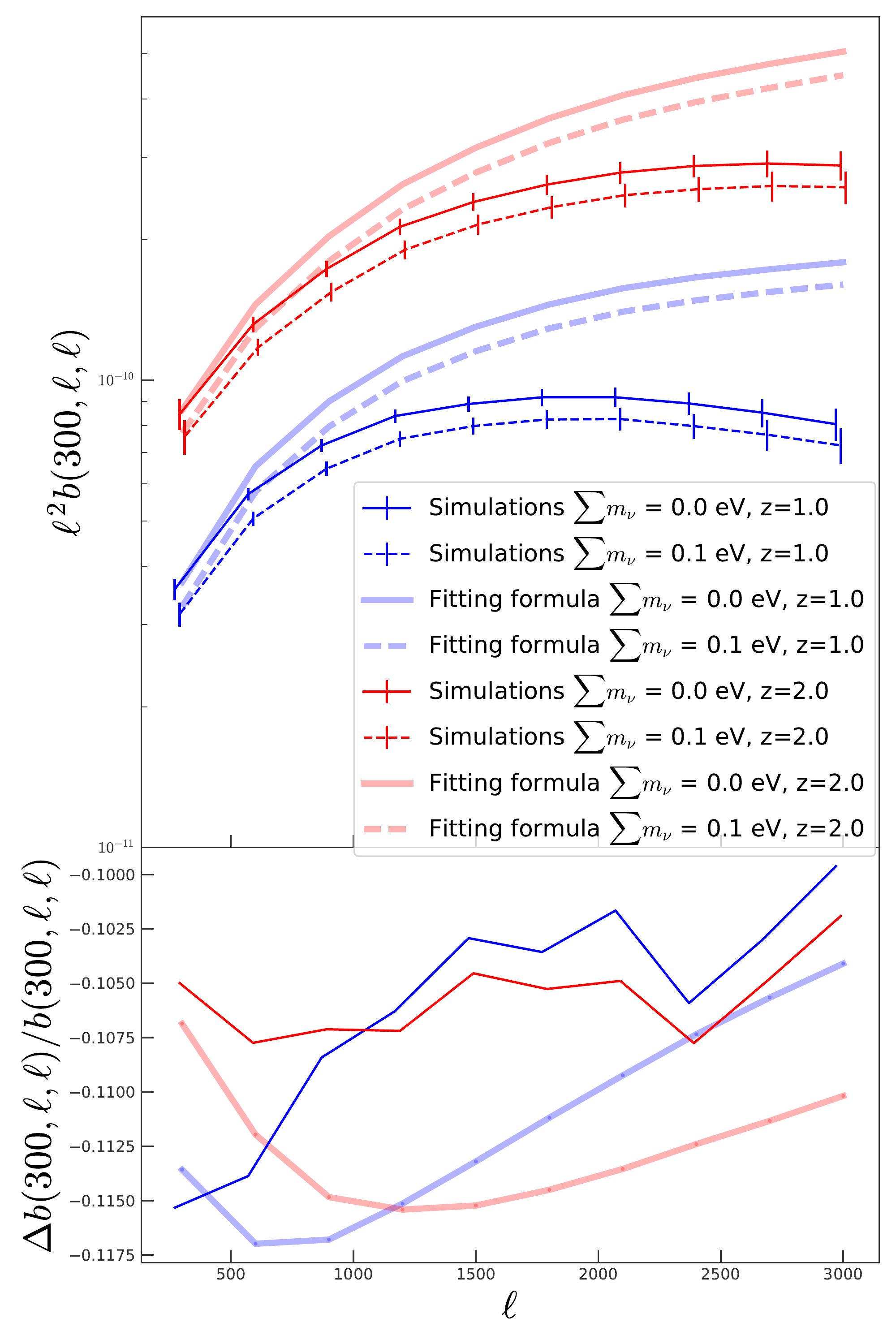}
    \label{fig:squeezeBispectrumSlice}}
    \qquad
\subfloat[Equilateral Configuration]{
    \centering
    \includegraphics[width=.45\textwidth]{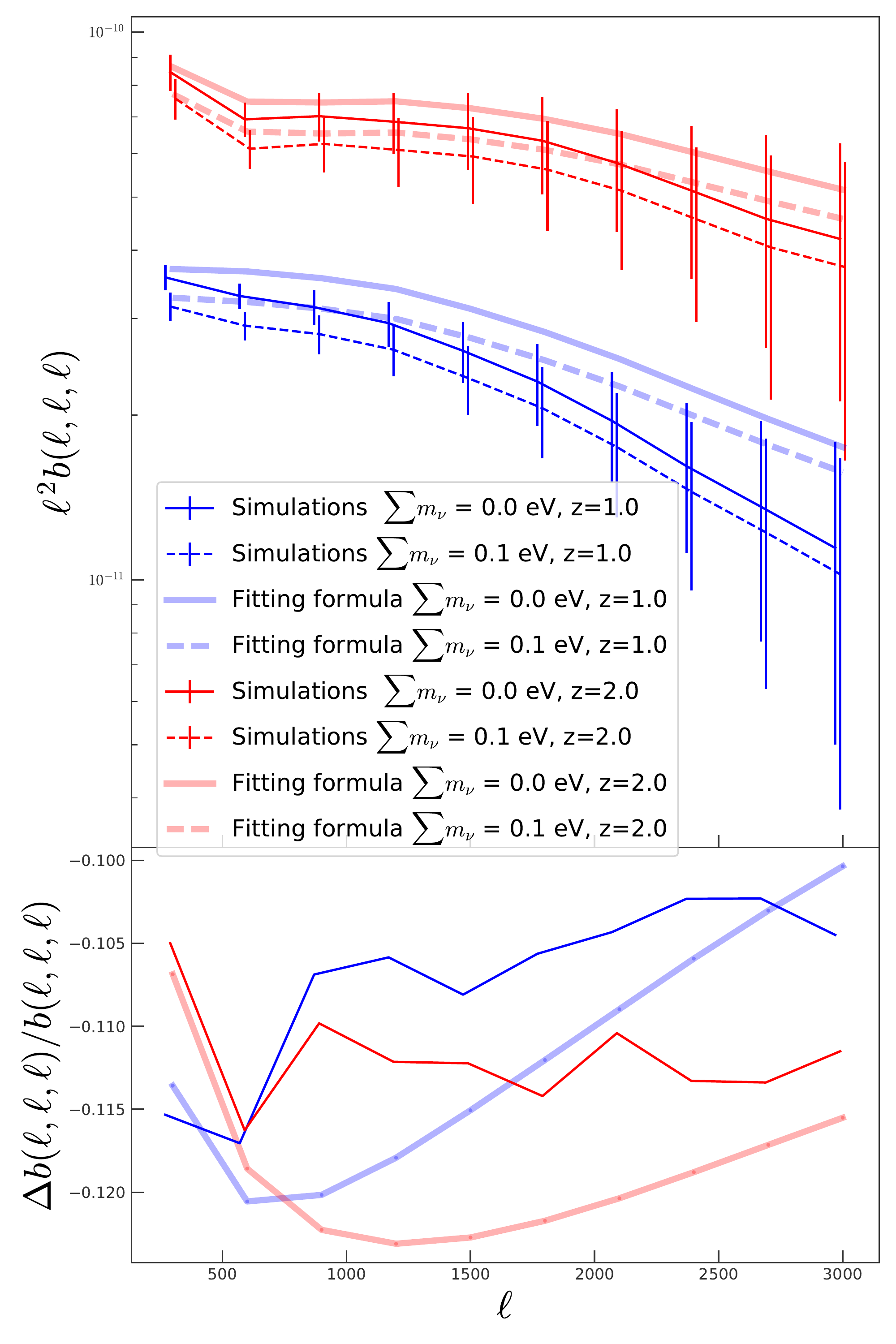}
    \label{fig:equilateralBispectrumSlice}}
\caption{In Figures \ref{fig:squeezeBispectrumSlice} and \ref{fig:equilateralBispectrumSlice} the squeezed and equilateral slices for the massless ($\sum m_\nu=0.0$~eV, $\Omega_m=0.3$, $A_s=2.1\times 10^{-9}$) and massive ($\sum m_\nu=0.1$~eV, $\Omega_m=0.3$, $A_s=2.1\times 10^{-9}$) models are plotted, for $z=1$ and $z=2$. The bottom panels show the fractional difference between the two cases. The error bars show how well these configurations can be measured with an LSST-like experiment that observes half of the sky. We also overplot the fitting formula from \citet{Gil-Marin2012AnBispectrum}. The  disagreement between the simulations and the fitting formula at small scales arises due to the finite resolution of the simulations, though the relative differences (bottom panels) are consistent within the expected uncertainties of the fitting formula.}
\end{figure}

In figures \ref{fig:squeezeBispectrumSlice}  and \ref{fig:equilateralBispectrumSlice} we show the effect of neutrino mass on the equilateral and squeezed slices of the shear bispectrum. The most significant effect of the massive neutrinos on the shear-bispectrum is a suppression of the amplitude. We find that, compared to the zero neutrino mass case, the amplitude of most configurations is reduced by $~10 \%$ for $\sum m_\nu=0.1$~eV. \citet{Levi2016MassiveTheory} computed the effect of neutrino mass on the matter bispectrum to second order in perturbation theory. They found on small scales the equilateral bispectrum was suppressed by $-13.5 \Omega_\nu/\Omega_m$, which for $\sum m_\nu=0.1$eV is a suppression of $9.9\%$, and is consistent with the suppression seen in Fig. \ref{fig:equilateralBispectrumSlice}. \citet{Ruggeri2018} examined the effect of neutrino mass on the matter bispectrum with simulations and found, for $\sum m_\nu=0.17$eV, the amplitude of the matter bispectrum was reduced by $~16 \%$ on the smallest scales, which is consistent with our results. Thus the amplitude of the bispectrum is almost twice as sensitive to neutrino mass than the power-spectrum, which is reduced by $\sim 6\%$ for $\sum m_\nu=0.1$~eV. However, whilst the bispectrum is more sensitive to neutrino mass, the bispectrum is harder to measure (for LSST, the signal to noise of the bispectrum is a factor of $\sim 5$ less than the power-spectrum). This is seen in figure \ref{fig:equilateralBispectrumSlice} where we show the expected LSST error bars. 

We find that, for both the squeezed and equilateral configuration, the effect of neutrinos is greatest on the largest scales, but we find only a very weak dependence on configuration. However, it should be noted that our ability to study the configuration dependence of the bispectrum is limited due to the size of simulated maps. The small patch size limits our access to large scale modes $\ell<150$ and hence extremely squeezed shapes. This also necessitates the use of large bispectrum bins, $\Delta \ell =300$, which potentially wash out structure in the bispectrum and is potentially suboptimal (as our weighting of different modes within the bin is suboptimal). We leave the investigation of these issues to future work.

We find that neutrino mass affects the bispectrum differently at different redshifts. This is seen also in Figures \ref{fig:squeezeBispectrumSlice} and \ref{fig:equilateralBispectrumSlice}  where the squeezed and equilateral slices of the bispectrum are plotted for two different redshifts. We find that for high redshift sources the main effect of neutrino mass is to reduce the bispectrum amplitude. Whereas for lower redshift sources we find a weak scale dependence. The importance of this redshift information is discussed in Section \ref{sec:constraints}.

In order to validate these results, we compare the simulation bispectra to results obtained from the \citet{Gil-Marin2012AnBispectrum} fitting formula, including the post-Born effects as described in \citet{Pratten2016ImpactCMB}. \citet{Gil-Marin2012AnBispectrum} provide a fitting formula for the matter bispectrum that we weight with the lensing kernels to obtain the lensing bispectrum. The fitting formula has the functional form of
\begin{equation}\label{eq:fittingFormula}
B(\bm{k}_1,\bm{k}_2,\bm{k}_3)=2 F^{\mathrm{eff}}(\bm{k}_1,\bm{k}_2,\Pi)P(\bm{k}_1)P(\bm{k}_2)+\textnormal{cyclic permutations},
\end{equation}
where $P(\bm{k})$ is the nonlinear matter power spectrum, $F^{\mathrm{eff}}(\bm{k}_1,\bm{k}_2,\Pi)$ is a kernel based on the tree level perturbation theory with modifications fitted to simulations, and $\Pi$ denotes that this kernel has some dependence on cosmological parameters. \citet{Pratten2016ImpactCMB} showed that for low redshift sources, post-Born corrections on the bispectrum are small, which agrees with our findings. For the fitting formula, we use the same bins as  our simulations and use the CLASS \citep{Blas2011TheSchemes} implementation of  the HALOFIT \citep{Takahashi2012} matter power spectrum model. The fitting formula results are plotted in \ref{fig:squeezeBispectrumSlice} and \ref{fig:equilateralBispectrumSlice}. We see that for the $z=2$ equilateral configuration we have very good agreement between the \citet{Gil-Marin2012AnBispectrum} fitting formula and the simulations. For the other configurations we find good agreement on the largest scales but significant differences on smaller scales. These differences are driven by the resolution limitations of the simulation and are similar to the effects seen in the power spectrum (Fig. \ref{fig:powerspectrumResponse}). The finite mass resolution of the simulations results in a suppression of small scale structure and thus a suppression of bispectra configurations that probe the small scales. This effect is stronger for lower redshifts as an object at the resolution limit will subtend a larger angle at lower redshifts. The comparison to the fitting formula provides an important cross check of our simulations, however it has its own limitations as the fitting formula is only accurate to the $\sim 10\%$ level, 
for $0.03<k<0.4  h /\mathrm{Mpc}$ and $0\leq z \leq 1.5$.

In the lower panels of Figures \ref{fig:squeezeBispectrumSlice} and \ref{fig:equilateralBispectrumSlice} we compare the fractional effect of neutrino mass on the bispectrum. We find that the fitting formula shows slightly enhanced suppression. This arises as the matter power spectrum from CAMB with \citet{Takahashi2012} HALOFIT shows enhanced suppression (as seen in \citet{Liu2018MassiveNuS:Simulations}) and this then propagates to the bispectrum. In Appendix \ref{app:powspecTests} we explore how different HALOFIT models, with different levels of neutrino suppression, affect power spectrum constraints. We find that our simulation power spectrum constraints are consistent with the uncertainty displayed with these different models. As the cosmological dependence of the bispectrum can be predominantly captured with the power spectrum dependence of the fitting formula (Eq. \ref{eq:fittingFormula}), we expect that the deviations seen in Figures \ref{fig:squeezeBispectrumSlice} and \ref{fig:equilateralBispectrumSlice} are consistent with the theoretical uncertainty (represented by the spread of results in different halo models). 

\section{Constraints}\label{sec:constraints}
\begin{figure}
  \centering
    \includegraphics[width=.80\textwidth]{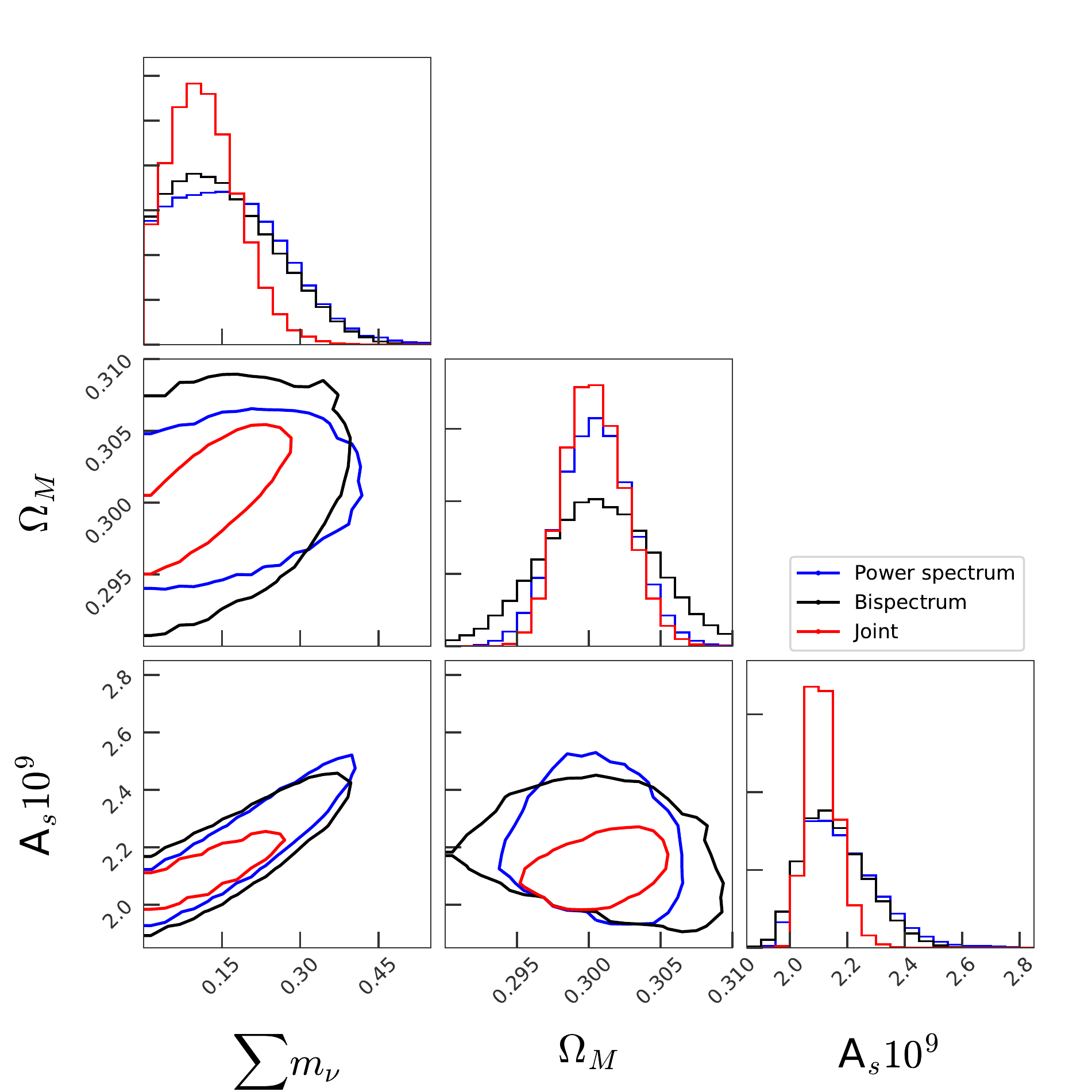}
\caption{ 95$\%$ parameter constraints from the power-spectrum, bispectrum and joint analysis. Here we set a fiducial model of $\sum m_\nu=0.1$~eV, $\Omega_m=0.3$ and $A_s \times 10^{9}=2.1$, and assume an LSST-like sky coverage, galaxy shape noise and number density. We also use five tomographic source redshifts. We find that the joint result improve upon the power spectrum parameter constraints by $\sim 32\%$  for $\sum m_\nu$, $13\%$ for $\Omega_m$ and $57\%$ for $A_s$. }
\label{fig:powspec_bi_joint_comp_noLens}
\end{figure}

In figure \ref{fig:powspec_bi_joint_comp_noLens} we present the constraints obtained from the bispectrum. We find a strong degeneracies between $\sum m_\nu$ and $A_s$, and between $\sum m_\nu$  and $\Omega_m$. This degeneracy arises as the shape of the bispectrum depends only weakly on the cosmological parameters. Increasing the neutrino mass or decreasing the matter density both reduce the amplitude, thus leading to a degeneracy between these parameters. If $\sum m_\nu=0.1$eV,  then LSST shear power spectrum would provide a $95\%$ constraint on the sum of the neutrino mass $\sum m_\nu<0.35$. 
The constraints from the bispectrum are similar to those from the power spectrum, for which the corresponding constraint is $\sum m_\nu<0.34$ 
We find that the bispectrum provides comparable constraints to the power spectrum across all three parameters, as shown in Fig. \ref{fig:powspec_bi_joint_comp_noLens}.

Despite the relatively similar degeneracies, we find that combining the power-spectrum and bispectrum measurements significantly improves the the constraints, as seen in  Fig. \ref{fig:powspec_bi_joint_comp_noLens}. We find that the joint measurements improves the constraint on $\sum m_\nu$ by 32$\%$, to 
$\sum m_\nu<0.23$, on $\Omega_m$ by $13\%$, 
and on $A_s$ by $57\%$. 

\begin{figure}
\centering
\includegraphics[width=.80\textwidth]{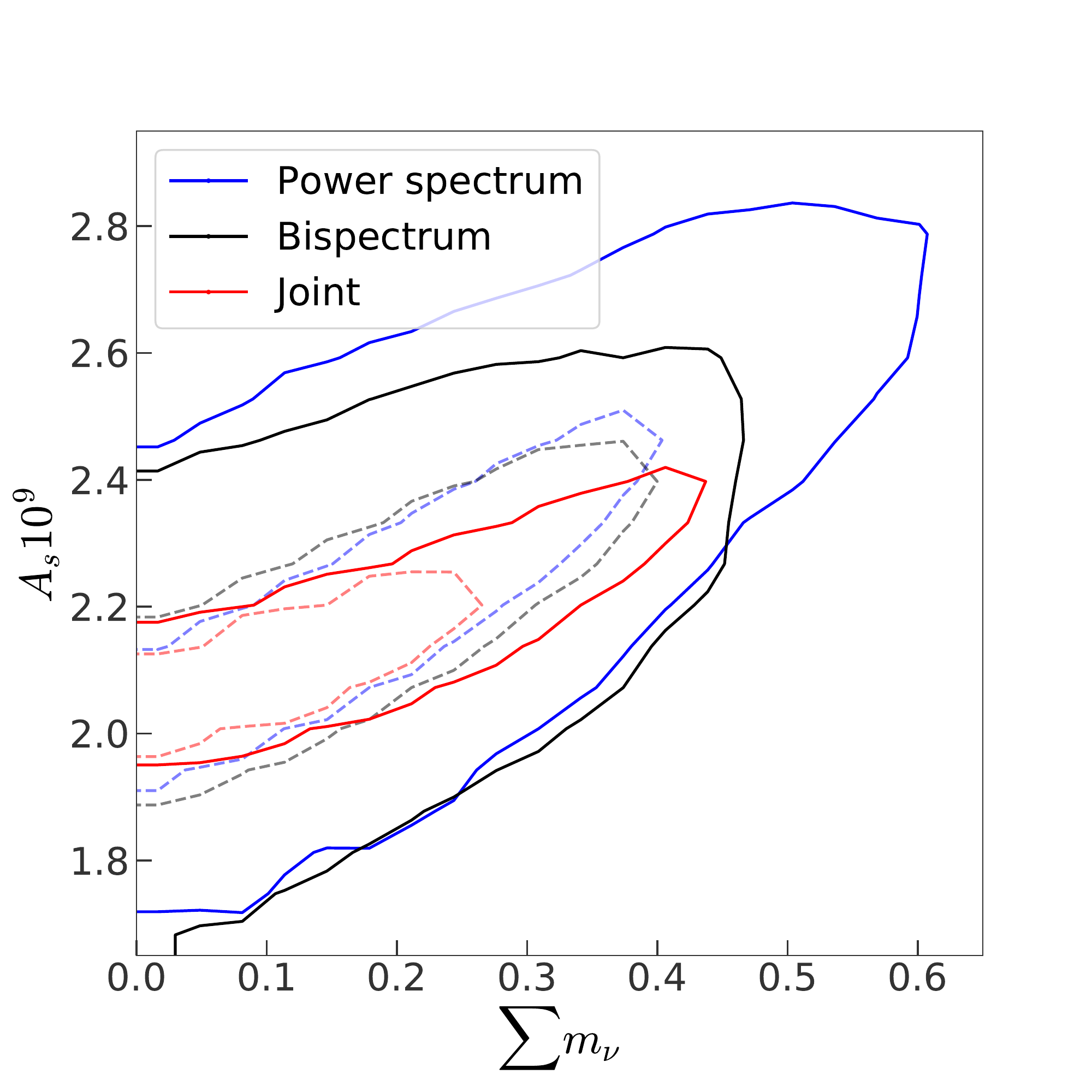}
\caption{ A comparison between power-spectrum and bispectrum $95\%$ constraints on $A_s $ and $\sum m_\nu$ from a single redshift measurement at $z=1.0$ against tomographic measurements from five redshift bins, $z\in \{0,0.5,1,1.5,2.,2.5\}$. The solid lines are the results from only a single redshift bin and the dashed lines are the results from tomography. The total number of galaxies per arcmin$^2$ is the same for the two cases.}
\label{fig:BispecTomography}
\end{figure}

\begin{figure}
\centering
\includegraphics[width=.8\textwidth]{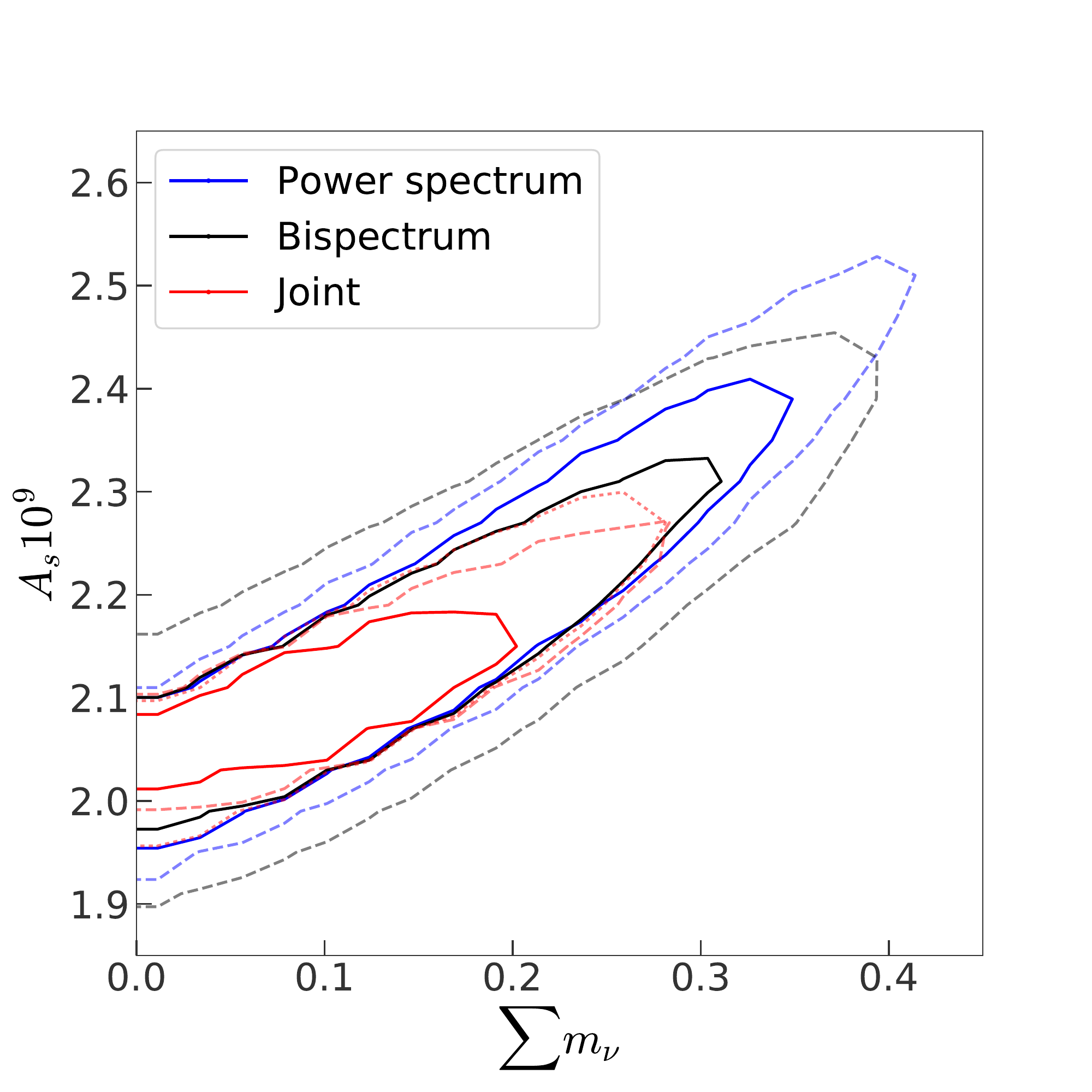}
\caption{A comparison of $95\%$ constraints on A$_s$ and $\sum m_\nu$ obtaining using a Gaussian covariance matrix (solid lines) and the covariance matrix from the simulations (dashed lines). The red dotted line is the constraint obtained when neglecting the cross correlations between the two and three point function in the covariance matrix. This demonstrates that for upcoming experiments non-Gaussian contributions to the covariance matrices cannot be neglected.}
\label{fig:simVsGausCovMat}
\end{figure}

In Fig. \ref{fig:BispecTomography}  we examine the effect of redshift tomography on neutrino mass constraints.  In the case without tomography we use the approximation that all of the galaxies lie with the redshift $z=1.0$ bin, thus it has $\bar{n}_{\rm{gal}}=44.8$ gal per arcmin$^2$. As was shown by \citet{Hu1999PowerLensing,Simon2004TheInformation}, tomography greatly enhances cosmological constraints with the power-spectrum. We see a similar result in Fig. \ref{fig:BispecTomography}. The growth information from tomography helps break the $A_s$ and $\sum m_\nu$ degeneracy, and hence significantly improves the power-spectrum constraint. For the bispectrum, we find that using tomography provides an improvement in the error on $A_s$ and $\sum m_\nu$, however it is less significant than the power-spectrum. The bispectrum is sensitive to the growth history so one would expect greater improvement with tomography than we see. However, as is seen in Figures \ref{fig:squeezeBispectrumSlice} and \ref{fig:equilateralBispectrumSlice} the weak redshift dependence induced by neutrino mass, combined with noise levels that are comparable to the signal means that tomography is less important. Note that combining the power spectrum and bispectrum without tomography produces results comparable to the power spectrum with tomography.

Finally we overview the impact of the two main covariance matrix components.  The covariance matrix can be split into two components: the Gaussian (also known as the disconnected) and non-Gaussian (connected) parts. The Gaussian component has no bin-to-bin correlations except between bins of the same $\ell$ at different redshifts and can be calculated via Wick's theorem.  The non-Gaussian contributions to the weak lensing covariance matrices arise as the convergence field is not Gaussian and these contributions have been extensively studied \citep[e.g][]{Schaan2014JointVariance,Takada2013PowerCovariance,Barreira2018AccurateSimulations}. As previous works have noted \citep{Kayo2013InformationMatrix,Sato2013ImpactEstimation} non-Gaussian contributions to the covariance matrix can degrade parameter constraints by $10-30\%$. In Fig. \ref{fig:simVsGausCovMat} we show the degradation in constraints on $A_s$ and $\sum m_\nu$ when non-Gaussian contributions to the covariance matrices are included. In agreement with the previous work, we find that our errors increased by $\sim 30 \%$ when non-Gaussian contributions are included. These contributions become more significant as noise levels are pushed to lower and lower levels \citep{Sato2013ImpactEstimation}. For the joint constraint, we also plot the results obtained using non-Gaussian covariance matrices but neglecting the two and three point cross term. We see that this has a negligible effect on the constraints.\footnote{In fact the constraint without the cross term is slightly larger than the constraint that includes it. This arises due to an effect known as beat-coupling \citep{Hamilton2006OnSimulations}, which is a component of the super sample variance component of the non-Gaussian covariance. Here we only outline how this effect can produce tighter constraints and refer the reader to \citet{Sefusatti2006CosmologyBispectrum}, who discuss this effect in detail. Beat-coupling arises as there are long wavelength fluctuations, at or above the size of the observed field, which affect the power spectrum and bispectrum coherently and these effectively add noise to the power spectrum and bispectrum. However by including the cross term between the bispectrum and power spectrum, a measurement of one of these observables can be used to remove this noise from the other and, as these modes are near or above the sample size, this subtraction removes little cosmological information.} This result emphasizes that the bispectrum is adding independent information.

\section{Conclusions}\label{sec:conclusions}
In this work, we explored the impact of neutrino mass on the weak lensing bispectrum. 
We confirm that there is rich information beyond the power spectrum, and that the constraints from the bispectrum alone are already comparable to that from the power spectrum. Combining the power spectrum and bispectrum measurements helps break the parameter degeneracies, therefore the two statistics are highly complementary. Analyzing them jointly produces significantly tighter parameter constraints, and in particular, by $\sim 32\%$  for $\sum m_\nu$, $13\%$ for $\Omega_m$ and $57\%$ for $A_s$, compared to using the power spectrum alone.


In this work, we have only implemented noise from galaxy number density and shape measurements. We do not consider the impact of systematics such as source-lens clustering \citep{Bernardeau1997TheStatistics,Hamana2001Source-lensConvergence}, intrinsic alignments \citep{Hirata2003ReconstructionPolarization,Crittenden2001SpininducedMeasurements}, baryonic effects \citep{Semboloni2011QuantifyingTomography}, photometric redshift biases and catastrophic redshift errors \citep{Mandelbaum2008PrecisionLensing,Bernstein2010CatastrophicRequirements,Hearin2010ATOMOGRAPHY}, and multiplicative biases \citep{Massey2013OriginsInstrumentation,Huterer2006SystematicSelf-calibration,Schaan2017LookingLensing}, for example. In addition, we have only considered the effect of three cosmological parameters, while keeping other parameters fixed. It is known that for the power spectrum, curvature, the Hubble parameter and a time-dependent dark energy equation of state are degenerate with the neutrino mass sum \citep{Font-Ribera2014,Benoit-Levy2012,Hamann2012,Mishra-Sharma2018NeutrinoExperiments}, though we note that many of these effects may be alleviated by the inclusion of external datasets, such as  primary CMB, CMB lensing, baryon acoustic oscillation and Lyman-alpha data. Whilst all of these effects will impact the constraints, it is expected that the relative contribution of the bispectrum to power spectrum will be similar. We defer a thorough examination of these effects to future work. 


\section{Acknowledgements}
We are grateful to Jo Dunkley, Zoltan Haiman, Zack Li, Emmanuel Schaan, Emanuele Castorina and Ben Wandelt  for useful discussions. This work is partly supported by an NSF Astronomy and Astrophysics Postdoctoral Fellowship (to JL) under award AST-1602663. We acknowledge NASA for its support of the WFIRST and Euclid programs.
We thank New Mexico State University (USA) and Instituto de Astrofisica de Andalucia CSIC (Spain) for hosting the Skies \& Universes site for cosmological simulation products.
This work used the Extreme Science and Engineering Discovery Environment (XSEDE), which is supported by NSF grant ACI-1053575. This research used resources of the National Energy Research Scientific Computing Center (NERSC), a U.S. Department of Energy Office of Science User Facility operated under Contract No. DE-AC02-05CH11231. The analysis is in part performed
at the TIGRESS high performance computer center at
Princeton University.
\appendix
\section{Power-spectrum robustness tests}\label{app:powspecTests}

\begin{figure}
  \centering
    \includegraphics[width=.60\textwidth]{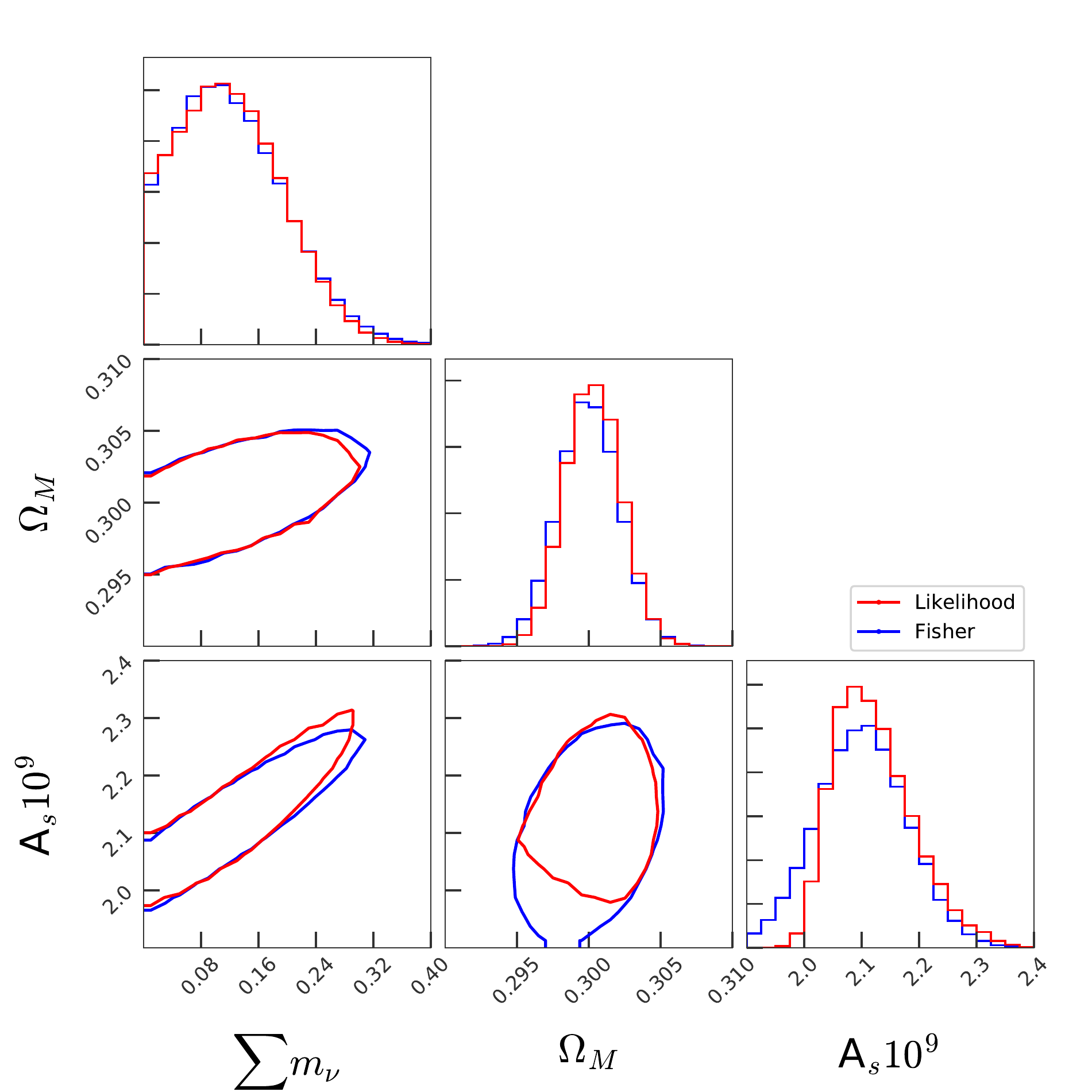}
\caption{A comparison between 95\% power spectrum constraints from a Fisher forecast with that from the interpolated-likelihood analysis, both use the HALOFIT model. The consistency between the Fisher forecast and our emulator provides a cross check of the accuracy of our emulator. The disagreement between the Fisher and emulator results arises as the Fisher constraint assumes that the constraints are symmetric Gaussians whereas the emulator explores the full posterior.}
    \label{fig:fisherVsEmulator}
\end{figure}
We performed a series of cross checks in order to validate both the performance of the interpolation and to compare our results to the commonly used HALOFIT predictions \citep{Takahashi2012}.

The first test compares the consistency of Fisher constraints with constraints from our interpolated-likelihood. We calculate the Fisher contours using the HALOFIT model with CAMB\citep{LewisCAMB}. For a fair comparison we then use CAMB to generate HALOFIT model power-spectra at the cosmologies of our simulations. We then interpolate these HALOFIT power-spectra in an identical manner to the simulations. We used a theoretical covariance matrix that only includes the Gaussian components. In Fig. \ref{fig:fisherVsEmulator} we plot the 95\% confidence contours from the HALOFIT Fisher forecast and from the HALOFIT interpolated-likelihood. In general we see good agreement between the two contours. The distributions are not identical; whilst the upper limits seem consistent, the lower limits are quite different. This arises as the Fisher constraint is symmetric and Gaussian whereas the interpolated constraint can be non-Gaussian.

\begin{figure}
  \centering
    \includegraphics[width=.60\textwidth]{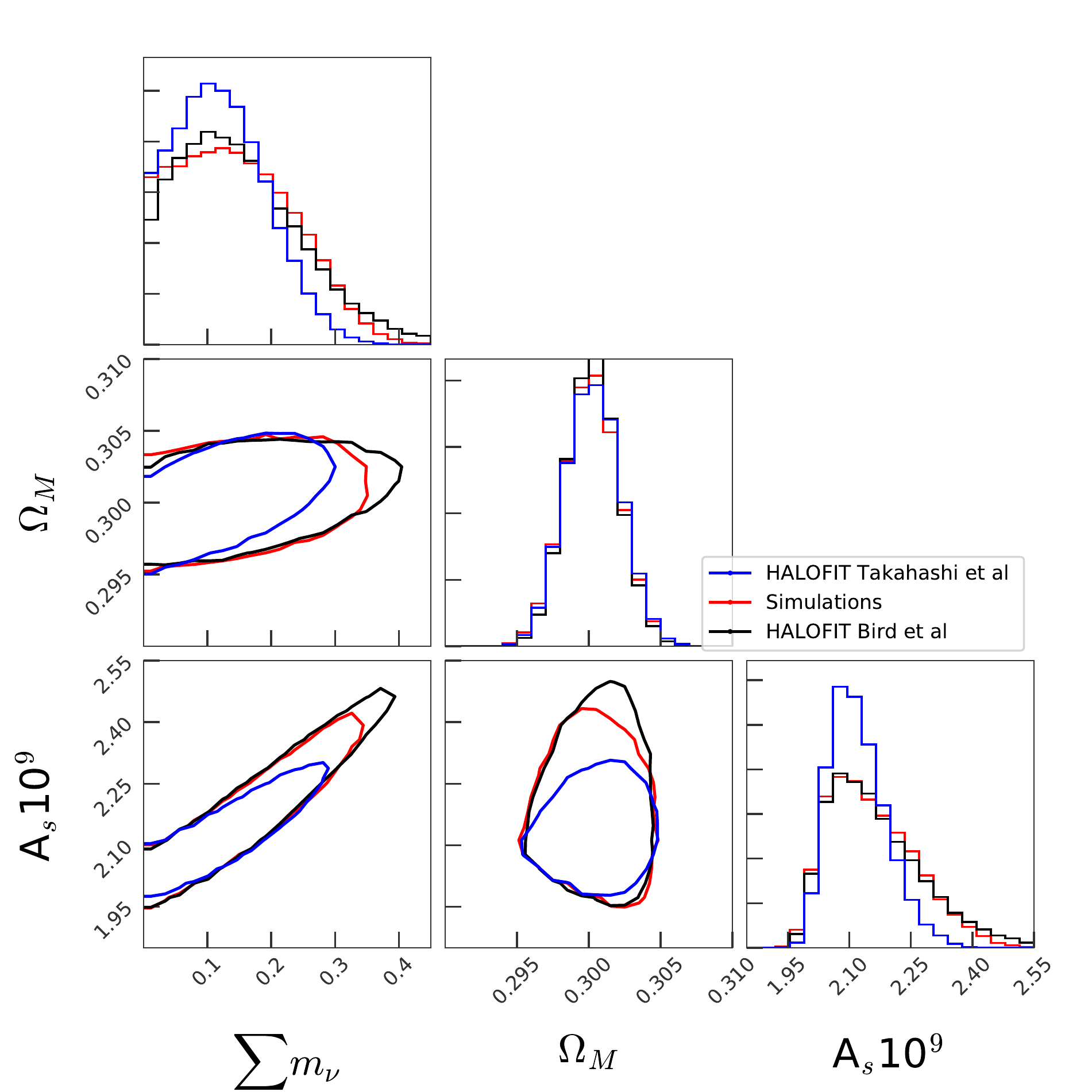}
\caption{A comparison between power spectrum constraints from two different HALOFIT models and our simulations. For the covariance matrix we use the theoretical Gaussian covariance matrix calculated separately for the HALOFIT models and simulation constraints. The contours are the $95\%$ confidence levels. The agreement between our simulations and CAMB demonstrates the robustness of our simulations.}
    \label{fig:powspecCAMBvsSims}
\end{figure}
The second test, shown in Fig. \ref{fig:powspecCAMBvsSims}, compares the constraints from the HALOFIT model with that from our simulations (again using theoretical Gaussian covariance matrices), and we use 
two different HALOFIT models from \citet{Takahashi2012} and \citet{Bird2012}. 
The Gaussian covariance matrices are calculated using their respective fiducial power spectrum for consistency.
These two HALOFIT models capture the theoretical uncertainty on the effect of massive neutrinos and the constraint from simulations lies between these models. This provides further validation of our simulations. The consistency between Fisher estimates and the emulator and between HALOFIT and the simulations provides confidence that our analysis pipeline is robust.

\bibliographystyle{act}
\bibliography{Mendeley,project}
\end{document}